\documentclass[twocolumn,preprintnumbers, amsmath, amssymb]{revtex4}
\def\be{\begin{equation}}
\def\ee{\end{equation}}
\def\beq{\begin{eqnarray}}
\def\eeq{\end{eqnarray}}
\def\n{\nonumber}
\def\bay{\begin{array}}
\def\eay{\end{array}}

\begin{document}
\preprint{CIRI/05-smw04}
\title{On the continuum origin of Heisenberg's indeterminacy
relations\footnote{This article is dedicated to Prof. Malcolm
MacCallum on his 60th birthday. His suave and helpful nature, his
easily accessible personality and his important scientific
contributions to general relativity have indeed been inspiring for
all the researchers of general relativity.}}

\author{Sanjay M. Wagh}
\affiliation{Central India Research Institute, \\ Post Box 606,
Laxminagar, Nagpur 440 022, India\\
E-mail:cirinag\underline{\phantom{n}}ngp@sancharnet.in}

\date{April 13, 2004}
\begin{abstract}
If space is indistinguishable from the extension of a physical
body, as is Descartes's conception, then transformations of space
become transformations of physical bodies. Every point of space
then has properties of physical bodies in some suitable
non-singular sense of average over the space. Every point of space
is then thinkable as a non-singular point particle possessing such
(averaged) physical properties. Then, the location of such a point
particle is, relative to another (similar) point particle, {\em
indeterminate\/} over the extension of the physical body. Further,
transformations of the space may ``move'' such a point particle in
relation to another such point particle. These notions then
provide a non-probabilistic explanation of Heisenberg's
indeterminacy relations.
\\

\centerline{To be submitted to: General Relativity \& Gravitation}
\end{abstract}
\maketitle

\newpage

\section{Introduction}
Newtonian mechanics ascribes independent and real existence to
space, time and matter. In Newton's theory, space and time also
play a dual role. Firstly, they play the role of a ``background''
for things happening physically to matter. Secondly, they also
provide us the inertial systems which happen to be advantageous to
describe Newton's law of inertia. Then, if matter were to be
removed completely, the space and time of the newtonian framework
would ``remain'' behind.

Descartes opposed \cite{ein-pop} the concept of space as being
independent of physical objects.  Essentially, he stated: the
space is identical with extension, but extension is connected with
physical bodies; thus there should be no space without physical
bodies and hence no empty space. There certainly are
\cite{ein-pop} (philosophical) weaknesses \footnote{The concept of
extension owes its origin to our experience of bringing into
contact physical bodies. But, from this alone it cannot be
concluded that the concept of an extension is unjustifiable in
cases which themselves have not given rise to the formation of
this concept.} of such an argument. However, let ``the space be
indistinguishable from physical bodies.''

The question is then of some suitable theoretical (mathematical)
formulation which describes space (and physical bodies) as per
Descartes's conception. Such a description can be expected to
possess the following characteristics.

In this description, Cartan's volume-form should be well-defined
at every spatial location. A point of space could also be
prescribed, in some suitable non-singular sense, the inertia,
electric charge etc. Then, we could also look at such a point as a
point particle in the newtonian sense. We may also look at a
physical body in the newtonian, non-singular, sense of a point
particle possessing various properties of a physical body. Thus,
physical bodies should {\em everywhere in space\/} be describable
as singularity-free.

In this description, there cannot therefore be {\em any\/} spatial
location without a physical property of a material object. Hence,
any local motion of a physical body will, clearly, be a change in
the (suitable) structure of the space.

The issue is of incorporating time in this framework. Now, the
temporal evolution of ``points of space'' is a mathematically
well-definable concept - as a dynamical system.

Then, Descartes's conceptions could be realizable in some
mathematically well-defined formalism that deals with dynamical
systems defined on continuum as the underlying set.

Such a description then also follows the principle of general
covariance: the laws regarding physical objects in it are based on
the arbitrary coordinate transformations of the underlying space
and also on time as an essentially arbitrary parameter of the
dynamical system.

The question now is of suitable mathematical structure on the
space that allows us the association of physical properties of
material objects to the points of the space. Furthermore, the
question is also of defining in a natural manner the boundary of
any physical object.

A physical object has associated with it various (fundamental)
physical properties, {\em eg}, (rest) energy. Then, the adjacent
objects clearly separate by boundaries at which the spatial
derivative(s) of that property under consideration, ({\em eg},
rest energy), change(s) the sign.

But, a physical object is a {\em region of space\/} as per
Descartes's conception. Therefore, some suitable structure
definable on the space must, as per Descartes's conception, then
possess a similar property of its derivative changing its sign at
a boundary of a physical object. It then also follows that a
physical property can, essentially, be specified {\em
independently\/} for each spatial direction since these directions
are to be treated as {\em independent\/} of each other.

Now, the space (continuum) is characterizable by ``distance''
separating its points. Suitable ``distance'' function can then be
expected \footnote{This is a ``necessary'' consequence of
Descartes's conception. We can, of course, choose the defining
property (of a physical object) itself as a ``distance''
function.} to possess the property of its derivative(s) changing
sign across boundaries separating regions of space corresponding
to separated objects. This suitable distance function then,
mathematically, becomes a pseudo-metric function on the space,
remaining a metric function within a region.

Consider therefore a three-dimensional pseudo-Riemannian manifold,
denoted as $\mathbb{B}$, admitting a pseudo-metric \cite{smw0}:
\setcounter{equation}{0}
\beq d\ell^2&=& {P'}^2Q^2R^2\, dx^2 \n \\
&\phantom{m}&\hspace{.1in} +\;P^2\bar{Q}^2 R^2\, dy^2 \n \\
&\phantom{m}&\hspace{.3in} +\;P^2Q^2\tilde{R}^2 \,dz^2
\label{3d-metric-gen} \eeq where we have $P\equiv P(x)$, $Q\equiv
Q(y)$, $R\equiv R(z)$ and $P'=dP/dx$, $\bar{Q}=dQ/dy$,
$\tilde{R}=dR/dz$. The vanishing of any of these spatial functions
is a {\em curvature singularity}, and constancy (over a range) is
a {\em degeneracy\/} of (\ref{3d-metric-gen}).

A choice of functions, say, $P_o$, $Q_o$, $R_o$ is a specific
distribution of ``physical properties'' in the space of
(\ref{3d-metric-gen}). As some ``region'' of physical properties
``moves'' in the space, we have the original set of functions
changing to the ``new'' set of corresponding functions, say,
$P_1$, $Q_1$, $R_1$.

Clearly, we are considering the isometries of
(\ref{3d-metric-gen}) while considering ``motion'' of this kind.
Then, we will remain within the group of the isometries of
(\ref{3d-metric-gen}) by restricting to the triplets of {\em
nowhere-vanishing\/} functions $P$, $Q$, $R$. We also do not
consider any degenerate situations for (\ref{3d-metric-gen}).

If we denote by $\ell$ the pseudo-metric function corresponding to
(\ref{3d-metric-gen}), then $(\mathbb{B},\ell)$ is an uncountable,
separable, complete pseudo-metric space. If we denote by $d$, a
metric function canonically \cite{kdjoshi} obtainable \footnote{We
define an equivalence relation ``$\sim$'' such that $x\sim y$ iff
$\ell(x,y)=0$ where $\ell$ is the pseudo-metric distance defined
on the space $X$. Denote by $Y$ the set of all equivalence classes
of $X$ under the equivalence relation $\sim$. If $A,B \in Y$ are
two equivalence classes, then let $e(A,B)=\ell(x,y)$ where $x\in
A$ and $y\in B$. The (metric) function $e$ on $Y$ is the canonical
distance.} from the pseudo-metric (\ref{3d-metric-gen}), then the
space $(\mathbb{B}, d)$ is an uncountable, separable, complete
metric space. If $\Gamma$ denotes the metric topology induced by
$d$ on $\mathbb{B}$, then $(\mathbb{B}, \Gamma)$ is a Polish
topological space. Further, we also obtain a Standard Borel Space
$(\mathbb{B},\mathcal{B})$ where $\mathcal{B}$ denotes the Borel
$\sigma$-algebra of the subsets of $\mathbb{B}$, the smallest one
containing all the open subsets of $(\mathbb{B},\Gamma)$
\cite{trim6}.

A measurable, one-one map of $\mathbb{B}$ onto itself is a Borel
automorphism. Now, the Borel automorphisms of
$(\mathbb{B},\mathcal{B})$, forming a group, are natural for us to
consider here.

But, the pseudo-metric (\ref{3d-metric-gen}) is a metric function
on certain ``open'' sets, to be called the P-sets, of its Polish
topology $\Gamma$. A P-set of $(\mathbb{B},d)$ is therefore never
a singleton subset, $\{ \{x\}:x\in \mathbb{B}\}$, of the space
$\mathbb{B}$. Note also that every open set of $(\mathbb{B},
\Gamma)$ is {\em not\/} a P-set of $(\mathbb{B}, d)$.

Now, the differential of the volume-measure on $\mathbb{B}$,
defined by (\ref{3d-metric-gen}), is \be d\mu \;=\;
P^2Q^2R^2\,\left( \frac{dP}{dx}\frac{dQ}{dy}
\frac{dR}{dz}\right)\;dx\,dy\,dz \label{volume1} \ee This
differential of the volume-measure vanishes when any of the
derivatives, of $P$, $Q$, $R$ with respect to their arguments,
vanishes. (Functions $P$, $Q$, $R$ are non-vanishing over
$\mathbb{B}$.)

A P-set of the space $\mathbb{B}$ is then also thinkable as the
interior of a region of $\mathbb{B}$ for which the differential of
the volume-measure, (\ref{volume1}), is vanishing on its boundary
while it being non-vanishing at any of its interior points.

Any two P-sets, $P_i$ and $P_j$, $i,j\;\in\;\mathbb{N}$,
$i\,\neq\,j$, are, consequently, {\em pairwise disjoint sets\/} of
$\mathbb{B}$. Also, each P-set is, in its own right, an
uncountable, complete, separable, metric space.

Evidently, {\em a P-set is the mathematically simplest form of
``localized'' physical properties in the space $\mathbb{B}$\/} and
we call it a {\em physical particle}. This suggests that suitable
mathematical properties of a P-set can represent physical
properties.

We then recall that the Galilean concept of the (inertial) mass of
a physical body is that of the measure of its inertia. Therefore,
some appropriate measure definable for a P-set is the property of
inertia of a physical body, a P-set in question. So also should be
the case with the gravitational mass of a physical body. Such
should also be the case with other relevant properties of physical
bodies, for example, its electric charge.

Also, {\em signed\/} measures are definable on a P-set as well.
Signed measures then provide us the notion of the ``polarity'' of
certain properties. For example, a signed measure can provide the
polarity of an electric charge.

Thus, we associate with every attribute of a {\em physical body},
a suitable class of (Lebesgue) measures on such P-sets. Therefore,
a P-set is a {\em physical particle}, always an {\em extended
body}, since a P-set cannot be a singleton set of
$(\mathbb{B},d)$.

Therefore, various physical properties (measures) {\em change\/}
only when the region of space (P-set) changes. Thus, a region of
space (P-set) and physical properties (the measures on P-sets)
are, then, are amalgamated into one thing here. This union of the
space and the physical properties is then clearly perceptible
here.

Moreover, a given measure can be integrated over the underlying
P-set in question. The integration procedure is always a
well-defined one for obvious reasons.

The value of the integral provides then an ``averaged quantity
characteristic of a P-set'' under question. Of course, this is a
property of the entire P-set under consideration.

For example, let us define an almost-everywhere finite-valued
positive-definite measurable function, $\rho$, on $(\mathbb{B},
\mathcal{B})$. Let us call its class the energy density.
Integrating it over the volume of a P-set, the resultant quantity
can be called a {\em total mass}, $m_{_T}$, of that P-set under
consideration. The total mass, $m_{_T}$, is a property of that
entire P-set and, hence, of {\em every point\/} of that P-set.

(Clearly, to define the notions of ``gravitational mass'' and of
``inertial mass'' of a P-set, we need to consider the ``motion''
of a P-set and also an appropriate notion of the ``force'' acting
on that P-set. Since we are yet to define any of these associated
notions, we call the integrated energy density as, simply, the
total mass of the P-set.)

We note that every point of the P-set is then thinkable as having
these averaged properties of the P-set and, in this precise
non-singular sense, is thinkable as a {\em point-particle\/}
possessing those averaged properties. It is in this non-singular
sense that points of the space $\mathbb{B}$ are point particles in
the present framework.

Clearly, the ``location'' of the mass $m_{_T}$ will be {\em
indeterminate\/} over the {\em size\/} of that P-set because the
averaged property is also the property of every point of the set
under consideration.

Now, in a precise mathematical sense \cite{kdjoshi}, sets can be
touching and that describes our intuitive notion of touching
physical bodies. Of course, the corresponding point particles are
then ``touching'' within the limits defined by the sizes,
boundaries, of the corresponding P-sets.

A Borel automorphism of $(\mathbb{B},\mathcal{B})$ then induces an
associated transformation of $(\mathbb{B}, d)$, say, to
$(\mathbb{B}, d')$, and that ``moves'' P-sets about in
$\mathbb{B}$, since (suitably defined) distance between the P-sets
can change under that Borel automorphism.

The individuality of a point particle is clearly that of the
corresponding P-set. As noted before, in the present formalism, a
point particle is a point of the P-set of the space $\mathbb{B}$
with associated integrated measures defined on that P-set. As a
Borel automorphism of the space $\mathbb{B}$ changes that P-set,
the integrated properties also change and, hence, the initial
particle changes into another particle(s), since integrated
measures change.

Here, the notion of energy of a point particle is then that of
some suitably defined integrated measure defined on a P-set of the
space $\mathbb{B}$. The notion of the momentum of a point particle
(as a point of the P-set of $\mathbb{B}$ with associated
integrated measures on that P-set) is then that of the
appropriately defined notion of the rate of change of, some
suitably defined, ``physical distance'' under the action of a
Borel automorphism of $\mathbb{B}$, including evidently any
changes that may occur to measures definable on that P-set.
Therefore, the notions of energy and momentum of a particle are
certainly (well-) definable in the present formalism.

Of course, we then need to discover various laws of such
transformations of point particles into one another in the present
formalism. But, it is clear at the outset that these will
crucially depend on the structure of the group of Borel
automorphisms of the space $\mathbb{B}$.

Further, if a P-set splits into two or more P-sets, we have the
process of {\em creation of particles\/} since the measures are
now definable individually over the split parts, two or more
P-sets. On the other hand, if two or more P-sets unite to become a
single P-set, we have the process of {\em annihilation of
particles\/} since the measures are now definable over a single
P-set.

Clearly, the {\em laws of creation and annihilation of
particles\/} will require of us to specify the corresponding
transformations causing the splitting and the merger of the
P-sets.

Now, we call as {\em an object\/} a region of $\mathbb{B}$ bounded
by the vanishing of (\ref{volume1}) but containing interior points
for which it vanishes (so such a region is not a P-set). Such a
region of $\mathbb{B}$ is then a collection of P-sets. But, a
P-set is a particle. Therefore, an {\em object\/} is a {\em
collection\/} of particles.

Objects may also unite to become a single object or an object may
also split into two or more than two objects under transformations
of P-sets. We may then also think of the corresponding laws for
these processes involving objects.

Moreover, the metric of $(\mathbb{B}, d)$ allows us the precise
definition of the sizes of P-sets and objects. Then, given an
object of specific size, we may use it as a {\em measuring rod\/}
to measure ``distance'' between two other objects.

We call this the ``physical'' distance separating P-sets (as
extended bodies). We also (naturally) define distance separating
objects.

Now, the Borel automorphisms of the space $\mathbb{B}$ can be
classified as

\begin{description} \item{(1)} those which {\em preserve\/} and,
\item{(2)} those which {\em do not
preserve} \end{description} measures defined on a specific P-set.

Note that we are restricting our attention to only a specific
P-set/Object and not every P-set/Object is under consideration
here.

Measure-preserving Borel automorphisms of the space $\mathbb{B}$
then ``transform'' a P-set maintaining its characteristic classes
of (Lebesgue) measures, that is, its physical properties.

Non-measure-preserving Borel automorphisms change the
characteristic classes of Lebesgue measures (physical properties)
of a P-set while ``transforming'' it. Evidently, such
considerations also apply to objects.

It is therefore permissible that a particular {\em periodic Borel
automorphism\/} leads to an {\em oscillatory motion\/} of a P-set
or an object while preserving its class of characteristic
measures.

We can then think of an object undergoing periodic motion as a
(physically realizable) time-measuring clock. Such an object
undergoing oscillatory motion then ``measures'' the time-parameter
of the corresponding (periodic) Borel automorphism since the
period of the motion of such an object is precisely the period of
the corresponding Borel automorphism.

Then, within the present formalism, a {\em measuring clock\/} is
therefore any P-set or an object undergoing {\em periodic motion}.

Then, crucially, the present formalism represents measuring
apparatuses, measuring rods and measuring clocks, on par with
every other thing that the formalism intends to treat.

Such considerations then suggest an appropriate distance function,
{\em physical distance}, on the family of all P-sets/objects of
the space $(\mathbb{B},d)$. More than one such distance function
will be definable, depending obviously on the collection of P-sets
or objects that we may be considering in the form of a measuring
rod or measuring clock.

This above is permissible since we are dealing here with a
continuum which is a standard Borel space with Polish topology.
Relevant mathematical results can be found in \cite{trim6}.

A Borel automorphism of $(\mathbb{B},\mathcal{B})$ may change the
physical distance resulting into ``relative motion'' of objects.
We also note here that the sets invariant under the specific Borel
automorphism are characteristic of that automorphism. Hence, such
sets will then have their distance ``fixed'' under that Borel
automorphism and will be stationary relative to each other.

On a different note, an automorphism, keeping invariant a chain of
objects separating two other objects, can describe the situation
of two or more relatively stationary objects.

Effects of the Borel automorphisms of $(\mathbb{B},\mathcal{B})$
on the (mathematically well defined) physical distance are then
motions of physical bodies.

Obviously, various concepts such as the density of point
particles, a flux of point particles across some surface etc.\ are
then well definable in terms of the transformations of P-sets and
the effects of these transformations on the measures definable
over the P-sets under consideration.

Then, we note that such ``averaging procedures'' are well-defined
over any collection of P-sets and, also, of objects. Thus, we may,
in a mathematically meaningful way, define a suitable
``energy-momentum tensor'' \footnote{This is a field-theoretic
comprehension, in a definite sense, of the energy-momentum
tensor.} and some relation between the averaged quantities, an
``equation of state'' defining appropriately the ``state of the
fluid matter'' under consideration.

(Such conceptions require however the notion of transformations of
P-sets and objects. Moreover, this averaging is a ``sum total'' of
the effects of various such transformations of P-sets and objects
and, hence, will require corresponding mathematical machinery.
This is, then, the premise of the ergodic theory. Recall that
$(\mathbb{B},\mathcal{B})$ is a Standard Borel Space.

Einstein's field equations are then definable in the sense (only)
of these averages. Therefore, Einstein's equations are
``obtainable'' on the basis of the temporal evolution of point
particles, points of the 3-space $\mathbb{B}$. Descartes's
conceptions are then also realizable in the present formalism.)

Clearly, a joint manifestation of Borel automorphisms of the space
$(\mathbb{B}, \mathcal{B})$ and the association,  as a point
particle, of integrated measures definable on a P-set with the
points of a P-set is a candidate reason behind Heisenberg's
indeterminacy relations in the present continuum formulation
\footnote{Recall that Einstein, in {\it Albert Einstein:
Philosopher - Scientist}\, (Ed. P. A. Schlipp, La Salle: Open
Court Publishing Company - The Library of Living Philosophers,
Vol. VII: 1970), p. 666, regarded the correctness of Heisenberg's
indeterminacy relations as being ``finally demonstrated''. But, he
looked for some non-probabilistic explanation for them.} since
intrinsic indeterminacy exists here.

Intuitively, as well as in a mathematically precise sense, it can
be seen that as the size of the P-set gets smaller and smaller the
position of the point-particle (of integrated characteristics of a
P-set) is determinable more and more accurately. (But, recall that
a P-set is never a singleton subset of $\mathbb{B}$. So, complete
positional localization of a point particle is not permissible.)

Now, any experimental arrangement to determine a physical property
of a P-set is based on some specific ``arrangement'' of P-sets and
involves corresponding Borel automorphisms of $\mathbb{B}$
affecting those P-sets. For example, Heisenberg's microscope
attempting the determination of the location of an electron
involves the collision of a photon with an electron. It therefore
has an associated Borel automorphism producing the motion of a
specific P-set, a photon.

Although we have not specified the sense \footnote{It follows that
this ``sense'' is that of measures definable on a P-set. We
therefore need to ``isolate'' the classes of measures
corresponding to a photon and an electron here.} in which a P-set
can be a photon, it is clear that the Borel automorphism causing
its motion will also affect an electron as a P-set.

Thus, a P-set ``transforms'' as a result of our efforts to
``determine any of its characteristic measures'' since these
``efforts or experimental arrangements'' are also Borel
automorphisms, not necessarily the members of the class of Borel
automorphisms keeping  invariant that P-set (as well as the class
of its characteristic measures).

Hence, a Borel automorphism (experimental arrangement)
``determining'' a characteristic measure of a P-set changes, in
effect, the very quantity that it is trying to determine. This
peculiarity then leads to Heisenberg's (corresponding)
indeterminacy relation.

Then, in the present continuum description, it is indeed possible
to explain the origin of Heisenberg's indeterminacy relations. The
present continuum description provides us therefore an origin of
indeterminacy relations. This is in complete contrast to their
probabilistic origin as advocated by the standard quantum theory.

Notice also that, in the present considerations, we began with
none of the fundamental considerations of the concept of a
quantum. But, one of the basic characteristics of the conception
of a quantum, Heisenberg's indeterminacy relation, emerges out of
the present formalism.

Then, in the present framework, we have also done away with the
``singular nature'' of the particles and, hence, also with the
unsatisfactory dualism of the field (space) and the source
particle. Furthermore, we have, simultaneously, well-defined laws
of motion (Borel automorphisms) for the field (space) and also for
the well-defined conception of a point particle (of integrated
measure characteristics). The present formalism is therefore a
complete field theory.

Further, none of the two notions of location and momentum is any
deficient for a description of the facts since Heisenberg's
indeterminacy relations are also ``explainable'' within the
present formalism. This explanation crucially hinges on the fact
that the points of the space $\mathbb{B}$, as singleton subsets of
the space $\mathbb{B}$, are never the P-sets. It is only in the
sense of associating the measures integrated over a P-set that the
points of the space $\mathbb{B}$ are point particles.

The measurable location of a particle is essentially a {\em
different\/} conception and that depends on the physical distance
definable on the class of all P-sets of the space $\mathbb{B}$.
The measurable momentum of a particle is also dependent on the
notion of the physical distance changing under the action of a
Borel automorphism of $\mathbb{B}$.

A comment on the mathematical methods would not be out of place
here. Then, we note that the mathematical formalism of the ergodic
theory is what is of immediate use for the present physical
framework. This much is already clear from the above
considerations.

However, it is not entirely satisfactory to use the present
methods of ergodic theory. One of the primary reasons for this
state of affairs is the inability of the present methods in
ergodic theory to let us handle, in a physical sense, the P-sets.
Some newer methods are then required here. 

\end{document}